\documentstyle[prl,aps,amssymb,epsf]{revtex}

\begin{document}

\title{Metastable States, Relaxation Times and Free-energy Barriers in
  Finite Dimensional Glassy Systems}

\author{Silvio Franz
}

\address
{ 
The Abdus Salam International Centre for Theoretical Physics,\\
Strada Costiera 11, P.O. Box 586, I-34100 Trieste, Italy\\ 
}


\maketitle

\begin{abstract}
  In this note we discuss metastability in a long-but-finite range
  disordered model for the glass transition. We show that relaxation
  is dominated by configuration belonging to metastable states and
  associate an in principle computable free-energy barrier to the
  equilibrium relaxation time. Adam-Gibbs-like relaxation times appear
  naturally in this approach. 
\end{abstract}


\section{Introduction}
Metastable states and configurational entropy, are central concepts in
our understanding of low temperature relaxation of disordered and
glassy systems \cite{Wa}.  In presence of time scale separation
between ``fast'' and ``slow'' degrees of freedom relaxation is often
described as a walk between metastable states, intended as regions of
configuration space where the fast degrees of freedom reach a state of
quasi equilibrium before the relaxation of slow degrees of freedom can
effectively take place.  According to the Adam and Gibbs theory
\cite{AG} the configurational entropy, i.e. the logarithm of the
multiplicity of metastable states is related to the relaxation time by
an inverse proportionality relation.

The relation between metastability and relaxation is
well understood in 
infinite range spin glasses with ``random first order transition''
\cite{W1}.  The dynamics of these systems is exactly described by the
equations of Ideal Mode Coupling Theory \cite{MCT}.  The Mode Coupling
transition, spurious in liquid systems, is here driven by the
existence of metastable states capable to trap the system for time
scales diverging with the systems size.  The divergence is of course a
Mean Field artifact, but features of mean-field metastable states
could well have a reflex in finite dimensional glassy systems
\cite{gold}. 

A well known problem in the study of metastable states is that while
they are easy to define within mean field theory, their definition
becomes elusive in finite dimensions.  This problem has been studied
in large detail in the context of systems with first order phase
transition, where the problem of metastability, can be considered as
satisfactorily understood \cite{L1,olivieri}.  Unfortunately, this is
not the case in glassy systems, where despite the appeal of a
description of dynamics in terms of metastable states, only very rare
contributions have tried to clarify the meaning of metastability
beyond the phenomenological level \cite{BK}.

The ``random first order transition scenario'' has led to
phenomenological attempts to treat the problem of glassy relaxation,
and ergodicity recovery in analogy with kinetics of ordinary first
order phase transition \cite{W2,P1,BB}.  Liquids at low temperature
would appear as ``mosaic states'', in which locally frozen variables'
configurations could be composed combinatorially on a large scale. The
typical size of the rigid regions could be computed in a nucleation
theory, with the configurational entropy playing the role of an
ergodicity restoration bulk driving force, competing with some
postulated interface tension.  These developments stimulated first
principle calculations in microscopic disordered models. In ref.
\cite{io} and later in \cite{dsw} a statistical description of the
ergodic state and a computation of the glassy coherent length below
$T_c$ was proposed through the asymptotic analysis of a Landau-like
glassy effective free-energy derived from microscopic models.
Unfortunately, in that papers it was not possible to make an explicit
connection between the computed free-energy barrier and the relaxation
time of the system.  Scope of this letter is to discuss this
connection.  In order to do that, we start from an analysis of glassy
relaxation based on separation of time scales.  We argue that finite
dimensional relaxation is dominated by metastable states that can be
characterized along the classical lines of Lebowitz and Penrose (LP)
\cite{lp2}, first proposed to describe metastable phases of matter in
presence of first order phase transitions.  According to LP,
metastable states can be considered as constrained equilibrium
ensembles with: 1) an order parameter is homogeneous on a suitable
mesoscopic length scale 2) a large time life and 3) a very small
probability of return once abandoned. In analogy with the work of LP,
we use in our analysis models with long-but-finite range Kac kind of
interactions, which in our case have a disordered character.  These
offer the possibility of studying finite dimensional effects in an
expansion around mean field, and the local mean-field character of
correlations, postulated in the phenomenological mosaic description,
appears as a consequence of the range of interaction \cite{FT1}.

\section{The model}
Let us consider a spherical Kac p-spin glass
model\cite{io} defined for real spins $\sigma_i$ on the $d$-dimensional
hypercubic lattice $\Lambda$ of linear size $L$ and Hamiltonian
\begin{equation}
\label{hp}
H(\sigma,J)=
-\sum_{i_1,\cdots, i_{p}\in\Lambda }
J_{i_1\cdots i_{p}}
\sigma_{i_1}\cdots \sigma_{i_{p}}
\end{equation}
where the couplings $J_{i_1\cdots i_{p}}$ are i.i.d. Gaussian
variables with zero average and variance 
\begin{eqnarray}
\label{wij}
E(J_{i_1\cdots i_{p}}^2)=
\frac {1}{ 2} \gamma^{pd}
\sum_{k\in \Lambda}\psi(\gamma|i_1-k|)\cdots
\psi(\gamma|i_p-k|)
\end{eqnarray}
where $p$ is an integer $p\geq 3$ and $\psi(|x|)$, is a non-negative
integrable function verifying the normalization $\int d^dx\,
\psi(|x|)=1$.  With this choice, the couplings $J_{i_1\cdots i_{p}}$
are sensibly different from zero only if all pairs of variables
$|i_l-i_m| {<\atop\sim}\; \gamma^{-1}$ $\forall l\ne m=1,...,p$, 
so that only variables that are at
distances $|i-j]\; {<\atop\sim}\; \gamma^{-1}$ effectively interact.
The effective interaction range $\gamma^{-1}$ will be assumed to be
large throughout the paper. We partition the lattice in boxes $B_x$ of
a coarse graining length size $\ell$ and impose a local spherical
constraint $\sum_{i\in B_x}\sigma_i^2=\ell^d$ for all $x$. We are
interested to the regime where the three defining lengths are
supposed, as in LP, to verify the relation
\begin{eqnarray}
\log L << \ell <<\gamma^{-1}<<L
\label{regime}
\end{eqnarray}
and for definiteness we will have in mind the situation where
$\ell=\gamma^{-\delta}$ for some $\delta\in (0,1)$.  The model is
chosen in such a way to reduce to the usual mean-field spherical
$p$-spin model in the regime $\gamma L \approx 1$.  Let us recall the
physics of this case \cite{Abarrat} that will be useful in the finite
$\gamma$ case.  On lowering the temperature from the paramagnetic
region one encounters two transitions.  There is a first transition at
the Mode Coupling temperature $T_c$ where ergodicity is broken.  Below
$T_c$ an exponential number of ergodic components ${\cal N}\approx
e^{N\Sigma(T)}$ dominates the thermodynamics.  The configurational
entropy $\Sigma(T)$ decreases as a function of the temperature, until
it becomes zero at a ``Kauzmann temperature'' $T_K<T_c$, where a
second transition is met. 
\begin{figure}
\begin{center}
\epsfxsize=8cm
\epsffile{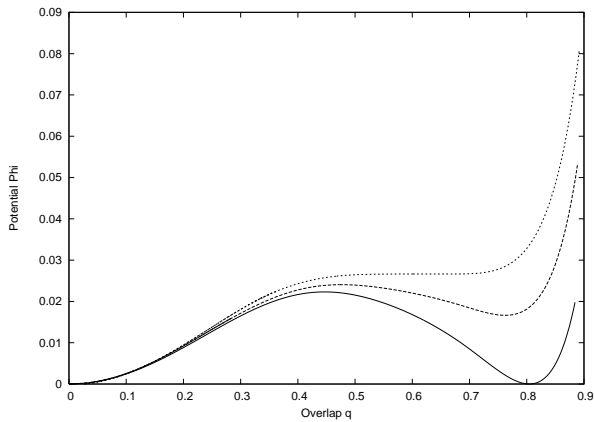} 
\end{center}
  \caption[0]{Qualitative behavior of the mean field potential. From
    top to bottom the $T=T_c$, $T_K<T<T_c$ and $T=T_K$. The difference
    of free-energy between the two minima is the product of the
    configurational entropy ad the temperature.
\label{pot}
}
\end{figure}
 This structure of states can be studied through the
so-called ``quenched glassy potential'' that considers the free-energy
$\Phi(q)$ of a system $\sigma$ constrained to have a fixed global
overlap $q=\frac 1{L^d} \sum_i \sigma_i\sigma^*_i$ with a typical reference
equilibrium configuration $\sigma^*$ \cite{FP1}.  This function has
the shape depicted in figure \ref{pot}, in the region $T_K<T<T_c$
one finds a two minimum structure analogous to the free-energy as a
function of the order parameter for mean-field systems with a
first-order phase transition.  
The value of the overlap in the low
minimum and high minima represent respectively the typical overlap
among random configurations in different and the same ergodic
components.  Both minima are associated to the exponential
multiplicity of ergodic components, and their difference gives
directly the configurational entropy times the temperature
$\Phi(q_{EA})-\Phi(0)=T\Sigma(T)$. 

\section{Dynamics at finite $\gamma$.}
\begin{figure}
\begin{center}
\epsfxsize=8cm
\epsffile{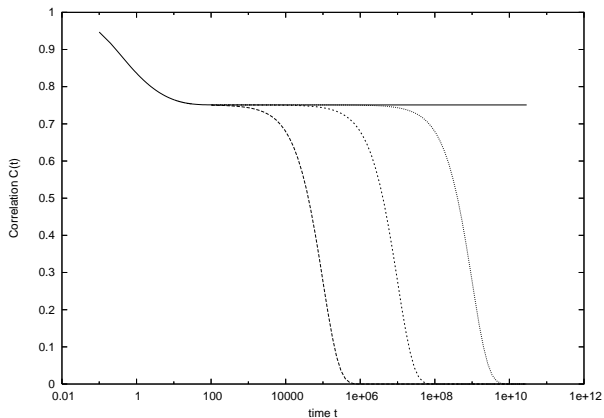} 
\end{center}
  \caption[0]{Qualitative behavior of the correlation function as a
    function of time for different values of $\gamma$. 
\label{fig1}}
\end{figure}

We now address the question of
restoration of ergodicity for finite $\gamma$ and $T_K<T<T_c$.
The relaxation of model (\ref{hp}) can be conveniently studied through
Langevin dynamics
\begin{eqnarray}
\frac{\partial \sigma_i}{\partial t}=\mu_x(t) \sigma_i-\frac{\partial
  H}{\partial \sigma_i}+\eta_i
\end{eqnarray}
where $\mu_x(t)$ are Lagrange multipliers that enforce in average the
local spherical constraint at each time and the $\eta_i$ are white
noise variables with correlations $E(\eta_i(t)\eta_j(s))=2T
\delta_{ij}\delta(t-s)$.  An insight on the dynamics in the regime
(\ref{regime}) can be obtained studying the Kac limit $\gamma\to 0$
after the thermodynamic limit. Through standard dynamical techniques
-Martin-Siggia-Rose formalism \cite{leticia} or dynamical cavity
method \cite{mpv}-, it is possible to conclude \cite{unp} that the finite
time local equilibrium spin-spin correlation function
$C_x^{(\gamma)}(t)=\frac 1{\ell^d} \sum_{i\in B_x}
\sigma_i(t)\sigma_i(0)$ on the scale $\ell$, tends to a space
homogeneous function, $C(t)$, that for temperatures larger then the
Kauzmann temperature $T_K$ verifies the usual equilibrium mean field
equation for the spherical p-spin model \cite{Abarrat}
\begin{eqnarray}
\frac{d C(t)}{dt}=-T C(t)+\frac{p \beta}{2} \int_0^t ds\;
C^{p-1}(t-s)\frac{d C(s)}{ds}. 
\end{eqnarray}
This equation predicts ergodicity breaking below $T_c$. The equilibrium
correlation function in this regime approaches exponentially fast on a
scale $\tau_{MF}$ a non zero limit $\lim_{t\to\infty}C(t)=q_{EA}$. 
Conversely, for any finite $\gamma>0$, for finite or infinite $L$, the
system should be able to escape from the metastable states and
$\lim_{t\to \infty} \lim_{L\to\infty} C_x^{(L,\gamma)}(t)$ coincide
with the equilibrium correlation function $\frac{1}{l^d} \sum_{i\in
  B_x} \langle \sigma_i\rangle^2=0$. In the regime (\ref{regime}), the
behavior of $C^{(\gamma)}_x(t)$ can be schematically depicted as in
figure \ref{fig1}.  As a function of time, the small $\gamma$ correlation
function remains close to the mean field curve up to some
characteristic time $\tau^{(\gamma)}$ diverging when $\gamma^{-1}\to
\infty$ and it decays to zero on longer scales.
Since for small enough $\gamma$ this characteristic time is arbitrary
larger than the mean field relaxation time $\tau_{MF}$, one finds that
the system equilibrates in the region of phase space where 
the local overlaps with the initial condition are close to $q_{EA}$
before relaxing further.  Moreover, one can expect that on time scales
of the order of $\tau_{MF}$, the system has homogeneous
correlations on the scale of the box size $\ell$, with typical
fluctuation from cell to cell of $C_x^{\gamma}(t)-C(t)$ scaling
as $1/\ell^{d/2}$. In turn, assuming independence of the fluctuations
beyond some finite length, using extreme statistics one can estimate
the maximum deviation $\max_x[C_x^{\gamma}(t)-C(t)]$ to be at
most of order $\sqrt{\frac{\log L}{\ell^d}}$. This, according to our
hypothesis, can be made arbitrarily small. The previously discussed
properties are enough to characterize  the
regions explored on scales smaller then $\tau^{(\gamma)}$ 
 as metastable states in the Lebowitz-Penrose sense.
The role of the ``order parameter'', homogeneous in space, is
played by the overlap of the configuration of the system with the
initial condition $\sigma(0)$.  Formally, for any typical equilibrium
configuration $\sigma^*$, we can define metastable states, as equilibrium
ensembles where the configurations $\sigma$ are such that in each box
$B_x$ the overlap with the initial configuration
$q_x(\sigma,\sigma^*)=\frac 1{\ell^d} \sum_{x\in B_x} \sigma_i\sigma^*_i$ is
suitably close to $q_{EA}$.  More precisely, one defines an interval
$I=[q_- ,1]$, with $q_- <q_{EA}$ and the set ${\cal R}
=\{\sigma|q_x(\sigma,\sigma^*)\in I\}$. The metastable state is the
restricted ensemble 
$$\mu_{{\cal R}}(\sigma)=\frac{1}{Z_{\cal R}} exp(-\beta
H(\sigma)){1}_{{\cal R}}(\sigma).$$
As in the LP case, the
definition of metastable state makes sense for a whole range of values
of $q_-$, such that a local fluctuation of the overlap larger then
$q_-$ is with high probability pushed back to $q_{EA}$. However, in
order for the definition to be useful, the value of $q_-$ should be
fixed a posteriori in such a way that a fluctuation larger then that
size should be amplified, and eventually run the system out of the
metastable state with finite probability.  We can now use this
analysis to estimate the relaxation rate of our system as a decay rate
of the metastable state.  We can then proceeds as in LP, and argue
that for the local Langevin dynamics we are using, the relaxation rate
$\lambda^{(\gamma)}=\frac 1{\tau^{(\gamma)}}$ is proportional to
$\lambda^{(\gamma)}\approx \frac{Z_{\partial{\cal R}}}{Z_{\cal R}}$
where $\partial {\cal R}$ is the set of configuration $\sigma$
belonging to ${\cal R}$, such that $q_{x_0}(\sigma,\sigma^*)=q_-$ for
at least one point $x_0$. $\lambda^{(\gamma)}$ is in principle a
random variable that depends on the initial configuration $\sigma^*$
and on the quenched random couplings $J_{i_1,...,i_p}$. However the
relaxation rate should be self-averaging with respect to both sources
of noise, and its typical value can be computed considering the
average of its log
\begin{eqnarray}
\log \lambda^{(\gamma)}_{typ}={  E}(\log Z_{\partial {\cal R}}-\log Z_{\cal R})
\label{quattro}
\end{eqnarray}
where ${E}$ stands for the canonical average over $\sigma^*$
and the quenched average over 
the distribution of the disorder $J$. Eq. (\ref{quattro}) is the main 
result of this paper. It connects the relaxation rate to a free-energy 
barrier computable in principle in purely static terms. 

\section{Relation with the Effective Potential.}
For fixed disorder and given finite volume we can define an 
 ``effective potential'' as the 
free-energy cost to keep the overlap with an equilibrium 
configuration $\sigma^*$ in each of the boxes $B_x$ fixed
to specified values $q_x$:
\begin{eqnarray}
\exp(-\frac \beta{\gamma^d} W_{\sigma^*,J}[q_x])= \sum_\sigma \prod_x
\delta(q_x(\sigma,\sigma^*)-q_x) e^{-\beta H(\sigma,J)}
\end{eqnarray}
where in the expression, we have anticipated the scaling of the
free-energy with the interaction range $\gamma^{-1}$ \cite{FT3}.
Is is clear how $Z_{\partial
{\cal R}}$ and $Z_{{\cal R}}$ can be related to $W_{\sigma^*,J}[q_x]$, in 
fact, for ${\cal S}={\cal R},\partial {\cal R}$, with transparent notation
\begin{eqnarray}
 Z_{\cal S} =\int_{\cal S} {\cal D}q_x e^{-
\frac \beta{\gamma^d}  W_{\sigma^*,J}[q_x]}. 
\label{sei}
\end{eqnarray}
By the previous discussion, the dominant profile in $Z_{{\cal R}}$ is
estimated as the constant $q_x=q_{EA}$ for all $x$.  For this profile,
the extensive part of $W_{\sigma^*,J}[q_x=q_{EA}]$ is a self-averaging
quantity: $V[q_{EA}]\approx{ E}(W_{\sigma^*,J}[q_x=q_{EA}])$.

For generic overlap profiles $q_x$, we can consider $V[q_{x}]={
E}(W_{\sigma^*,J}[q_x])$, and ask if the knowledge of this quantity is
enough to reconstruct the relaxation rate $\lambda^{(\gamma)}_{typ}$
 for small $\gamma$.

The restricted partition function $Z_{\partial {\cal R}}$ is:
$$
Z_{\partial {\cal R}}
= \sum_{x_0}
\int_{\{q_{x_0}=q_-; \; q_x\in I \; \forall x\}} {\cal D}q_x e^{-
\frac \beta{\gamma^d}  W_{\sigma^*,J}[q_x]}. 
$$ Each term of the sum on $x_0$ will be dominated by a single
profile, minimizing $W_{\sigma^*,J}[q_x]$ in presence of the
constraints. These will be profiles with a localized
excitation around $x_0$, corresponding to a finite free-energy cost
with respect to the uniform state $q_x=q_{EA}$. 

For fixed $\sigma^*$ and $J$, different values of $x_0$ could give
rise to different free-energy cost and different profile shapes.
However, we could argue again that on the mesoscopic scale
$\gamma^{-1}$ almost all sites $x_0$ would give the contributions of
the same order and these dominate the relaxation rate. This is a
stronger hypothesis than the other ones underlying our reasoning. It
implies that the incipient spatial heterogeneities on a scale
$\gamma^{-1}$ responsible for relaxation, have equal probability of
appearing in any point of space, independently of the reference
configuration $\sigma^*$ and the realization of the quenched disorder
$J$. Without that, the computation of
$\lambda_{typ}^{(\gamma)}$ should start from an evaluation of
(\ref{quattro}). This is in principle possible and when
done could allow the validation or refutation of the hypothesis. 

Previous work \cite{P1,io,dsw} was implicitly based on that strong
hypotheses. Within that framework, if we denote by
$q^*_x$ the profile maximizing $V[q_x]$ for $x_0=0$, the relaxation
rate will be given by
$$
\lambda^{(\gamma)}_{typ}=
\left(\frac L \ell \right)^d e^{-
\frac \beta{\gamma^d}  \left(V[q_x^*]-V[q_x=q_{EA}]\right)}
$$ and $V[q_x^*]$ will provide a Landau free-energy functional that we
can use in a nucleation-like theory. It is clear that restricted to
profiles constant in space, $q_x=q$, for small $\gamma$ we have:
$V(q_x=q)=W_{\sigma^*,J}[q_x=q]\approx(L\gamma)^d \Phi(q)$, where
$\Phi$ is the mean-field potential.  In order to compute the
free-energy barrier within this self-averaging setting, one should
consider spherically symmetric solutions with boundary conditions
$\lim_{|x|\to\infty}q_x=q_{EA}$.  Formulated in this way the problem
reduces to the computation of a critical droplet as suggested by
phenomenological theories \cite{W2,BB}.  According to conventional
nucleation theory \cite{L1}, close to $T_K$ where the two minima are
close to degeneracy, one can use the ``thin wall approximation'' and
evaluate the interface tension through the 1D instantonic solutions
connecting the two minima.  This competes with a bulk driving term
given by the free-energy difference between the minima, which in our
case is the mean-field configurational entropy. This first principle
formulation confirms the prediction of phenomenological theories, that
the configurational entropy provides the bulk driving force to
ergodicity restoration.  The actual computation of $V$ for fixed
profile can be performed through a modified version of the replica
method \cite{FP1}, needed to average over the spin reference
configuration $\sigma^*$ and the quenched disorder $J$.  This has been
done for the model (\ref{hp},\ref{wij}) in ref. \cite{io}.  Without
entering in the details of the discussion of that paper we just
mention that within a specific replica ansatz it is possible to find a
finite {\it surface} tension is found for $T$ approaching $T_K$. This
implies a glassy coherent length behaving as $R_g\sim \Sigma(T)^{-1}$
and an Adam-Gibbs free energy barrier $\Delta V \sim
\frac{1}{\gamma^d}\frac{C}{\Sigma^{d-1}}$. This result is supported by
a more elaborated ansatz in \cite{dsw}. As discussed in \cite{io}, and
in view of possible non self-averaging effects, these solution could
in fact just provide approximations to the true barrier. The ``right''
replica solutions could change the exponent $d-1$ appearing in the
relation. However, the general Adam-Gibbs structure of the formula,
that only depends on the fact that the bulk ergodicity restoring force
is the configurational entropy, should remain unchanged.

\section{Conclusions}
Summarizing, we have presented an argument that identifies as
Lebowitz-Penrose metastable states the dominant configurations of
relaxing glassy systems below the Mode Coupling transition. This
allows to compute the relaxation time in purely static term through
the evaluation of the probability of equilibrium fluctuations. The
relation between relaxation time and free-energy barrier is summarized
in eq. (\ref{quattro}) which is the main result of this letter.  We
have also discussed how previous results in the literature relate
implicitly to a self-averaging hypothesis. If this holds, the relevant
fluctuations have the shape of overlap droplets, and the bulk driving
force for ergodicity restoration is provided by the configurational
entropy. In agreement with phenomenological theories the relation of
the relaxation time with the configurational entropy follows a
generalized Adam-Gibbs equation.  The argument has been presented in
the case of disordered spin model, where playing with the interaction
range one can have time scales as much separated as wanted and a
mesoscopic scale for which correlations are homogeneous on short time
scales. On the other hand our argument is actually only based on the
hypothesis that time separation is such to allow partial equilibration
of the fast degrees of freedom for fixed values of the slow one, and
as such it should be generically valid when this condition is
respected.

{\bf  Acknowledgments}

I would like to thank M. M\'ezard and 
G.Parisi for discussions.  This work was supported in part by the
European Community's Human Potential programme under contract
``HPRN-CT-2002-00319 STIPCO''.

\end{document}